# Current Status and Perspectives of Cosmic Microwave Background Observations

Marco Bersanelli[(1,2)], Davide Maino[(1)], Aniello Mennella[(2)]

*(1) Dipartimento di Fisica, Università degli Studi di Milano, Via Celoria 16, 20133 Milano, Italy*
*(2) IASF-INAF, Sezione di Milano, Via Bassini 15, 20133 Milano, Italy*

**Abstract.** Measurements of the cosmic microwave background (CMB) radiation provide a unique opportunity for a direct study of the primordial cosmic plasma at redshift $z \sim 10^3$. The angular power spectra of temperature and polarisation fluctuations are powerful observational objectives as they encode information on fundamental cosmological parameters and on the physics of the early universe. A large number of increasingly ambitious balloon-borne and ground-based experiments have been carried out following the first detection of CMB anisotropies by COBE-DMR, probing the angular power spectrum up to high multipoles. The recent data from WMAP provide a new major step forward in measurements percision. The ESA mission "Planck Surveyor", to be launched in 2007, is the third-generation satellite devoted to CMB imaging. Planck is expected to extract the full cosmological information from temperature anisotropies and to open up new frontiers in the CMB field.

## INTRODUCTION

Observations of the cosmic microwave background (CMB) radiation probe the universe at an early stage of its evolution, corresponding to a redshift $z \sim 10^3$, characterised by a highly homogeneous plasma at temperature of ~3000 K with small fluctuations from uniformity of order $10^{-5}$. The CMB photons have travelled for 99.998% of the universe's age and provide us with a direct image of the primordial hot plasma, redshifted by cosmic expansion to millimeter wavelengths. A blackbody spectrum of the radiation is expected due to the thermal equilibrium of the primordial plasma. This prediction has been confirmed over a large frequency range, and measured with high-precision in the peak region by COBE-FIRAS [1,2]. The absence of significant distortions from a Planckian shape [3] testifies a smooth infancy of the universe and places upper limits to possible energy releases in the plasma era.

The first detection of temperature anisotoropies was obtained by COBE-DMR on scales >7° [4] and was later confirmed by sub-orbital experiments and refined by the full four-year DMR survey [5]. It is usual to represent the CMB temperature distribution on the celestial sphere by spherical harmonics expansion:

$$\Delta T(\theta,\phi) = \sum_{\ell,m} a_{\ell,m} Y_{\ell,m}(\theta,\phi)$$

where $\ell \approx 180°/\theta$, and $a_{\ell,m}$ represent the multipole moments that in simple models are characterized by zero mean, $<a_{\ell,m}> = 0$. The (non-zero) variance coefficients,

$C_\ell \equiv <|a_{\ell,m}|^2> \neq 0$, represent the CMB angular power spectrum (here the angle brackets indicate an average over all observers in the universe). The expected $C_\ell$ distribution is determined by acoustic peaks ([6] and references therein) whose scale and amplitude depend sensitively on the values taken by fundamental cosmological parameters such as the Hubble constant $H_0$; the spectral index of primordial fluctuations, $n_S$; the total energy density $\Omega_{tot}$ and its components such as the cold dark matter and barion density, $\Omega_{CDM}$ and $\Omega_B$, and the contribution from a cosmological constant, $\Omega_\Lambda$. For example, a first peak at $\ell \sim 200$ is expected for a flat $\Omega_{tot} = 1$ geometry. Accurate high-resolution imaging of the CMB over large portions of the sky represent therefore a powerful method to derive the value of such key parameters. The limited resolution of DMR allowed to cover the spectrum only up to $\ell < 20$. Here we summarise the current status of the observations (see [6] for more details).

## FROM COBE TO WMAP

COBE-DMR released the first anisotropy detection together with good quality full-sky maps, obtained with three pairs of Dicke-switched radiometers at 31.5, 53 and 90 GHz. The 4-year frequency-averaged map smoothed at 10° has a signal-to-noise ratio of ~2, thus providing a visual impression of the actual large scale CMB structure in the sky. The observed power spectrum, limited to $\ell < 20$, was found to be consistent with a scale-invariant spectral index ($n_S \sim 1$) of the primordial density fluctuations [7], and a power-law fit yielded an expected quadrupole term of $Q_{rms-PS} = 15.3$ μK [8]. As in any CMB experiment, the success of DMR depended on the accurate control of systematic effects [9] and calibration. The DMR results set the stage for new major efforts. The presence of temperature fluctuations at a measurable level $\Delta T/T_0 \sim 10^{-5}$ was demonstrated, and the race to unveil new features in the angular power spectrum had started. Increasingly ambitious sub-orbital programs were carried out mostly aiming at sub-degree scale detections in limited sky regions.

### Ground-Based Experiments

High altitude, dry sites have provided excellent results, in spite of the limitations of atmosphere and ground emissions. The Tenerife experiment, carried out at the Teide Observatory (2400 m), used similar technology as DMR in the 10-33 GHz range [10]. Clear detection of CMB structure was found at a level ~40 μK in the multipole range $\ell \sim 10$-30, overlapping with DMR. More recently, a 33GHz interferometer was installed [11] at Teide scanning a sky portion largely overlapping the original program. The detection of CMB anisotropy was unambiguous, with $\Delta T_\ell \sim 43$ μK at $\ell \sim 110$.

Even if harsh and isolated, such as the Antarctic Plateau, ground based sites offer the enormous advantage of reachable instruments and long integration times. The UCSB group lead to a series of degree-scale measurements from the Amundsen-Scott South Pole Station in the period 1988-1994 with HEMT receivers at 30-40 GHz, cooled at ~ 4K [12]. Significant correlated structure was observed with amplitude $\Delta T_\ell \sim 33$ μK at $\ell \sim 60$. The polar site also hosted the Python/Viper program, leading to detection at ~ 1° scales using an off-axis parabolic telescope that could be coupled either to bolometers at 90 GHz [13] or to a HEMT radiometer at 40 GHz [14]. The results were consistent with DMR at low $\ell$'s, and hinted an increase at $\ell > 40$. Viper was a follow up of Python at higher angular resolution [15] with observations at 40 GHz with a large 2.15 m telescope, yielding sensitivity in the range $100 < \ell < 600$.

With only a few exceptions [16], most ground based experiment take advantage of the atmospheric windows below 15 GHz, around 35 GHz and 90 GHz. The Saskatoon program produced one of the first convincing evidences of a raising spectrum (from $\Delta T_\ell \sim 49$ μK at $\ell \sim 87$ to $\Delta T_\ell \sim 85$ μK at $\ell \sim 237$). The instrument [17] used an off-axis parabolic reflector coupled to cryogenically cooled total power receivers based on low-noise HEMT amplifiers. As a follow-up, the MAT program probed smaller scales, $40 < \ell < 600$. The Saskatoon instrument was installed in a very high altitude site at Cerro Toco (5240 m) with the addition of a 144 GHz channel (0.2° beam), taking advantage of both HEMT and SIS technologies. The MAT/TOCO 31 and 42~GHz instruments were flown twice in 1996 (QMAP) to perform degree-scale observations providing detection in both flights [18].

At small angular scales, a series of filled-aperture observations at OVRO lead in the late 90's to an unambiguous detection [19] on scales ranging from 7' to 22'. Interferometers have provided high-quality observations at high $\ell$'s. The CAT three-element array operating at 13-17 GHz showed detection at $\ell \sim 420$ [20]. The VLA has been used to set limits to CMB anisotropy at sub-arcminute angular resolution since the early 80's and established new upper limits in the range 0.17' - 1.33' [21]. ATCA [22] yielded upper limits in polarised intensity and $\Delta T_\ell < 25$ μK for multipoles $3350 < \ell < 6050$. Finally, the SuZIE bolometer array [23] operated at Mauna Kea gave upper limits to CMB primary anisotropy. The combination of all data at high resolution gives evidence for a downturn in the power spectrum at sub-degree scales, as expected by standard cosmologies.

Three high quality ground-based data sets heve been obtained recently. The DASI 13-element interferometer [24] used cryogenic HEMTs in the 26-36 GHz spectral window and observed from the South Pole in the 2000 austral winter. The results mapped the power spectrum in the range $100 < \ell < 900$. The first peak at $\ell \sim 200$ was evident and in good agreement with results from other experiments. In addition, the DASI data suggested the presence of further peaks at $\ell \sim 550$ and $\ell \sim 800$. The CBI interferometer was sensitive to scales from 5' to 1° ($300 < \ell < 3000$). Excellent observations were obtained [25] with power spectrum results in good agreement with other experiments at $\ell < 2000$ but extending up to $\ell \sim 3000$. An excess power was observed at high $\ell$'s suggestive of a Sunyaev-Zeldovich effect was observed. Even more recently, the Viper 2.1 m telescope was used at the South Pole in the ACBAR experiment [26] to obtain two deep fields of the CMB, 3 degrees in size, with an rms of 8 μK per 5' beam.

## Balloon-Borne Experiments

Although the reduction of atmospheric emission from balloon altitudes (35-40 km) is great (factor $\sim 10^3$), in conventional flights the available observing time is only about 10-12 hours. In recent years, long-duration balloon (LDB) flights (10-15 days) have been successfully flown. Early balloon experiments confirmed the DMR detection.

The FIRS bolometers covered about 1/4 of the sky at ~3.8° resolution [27] in the range 170-680 GHz. The ARGO degree-scale bolometric experiment [28] lead to a detection of CMB anisotropy $\Delta T_\ell$ ~39 μK. BAM obtained statistically significant detection at degree scales ($\Delta T_\ell$ ~ 56 μK at $\ell$ ~ 75) [29] using a cryogenic differential Fourier transform spectrometer coupled to a 1.65 meter reflector. The MAX and MSAM balloon-borne multi-band bolometric receivers contributed to the progress at sub-degree scale. MAX was flown 5 times between 1989 and 1994 and observed in nine different sky regions yielding seven positive anisotropy detections [30]. The results also suggested a band power at degree-scales higher than DMR. MSAM [31] observed a 10 deg$^2$ region with 0.5° resolution. MSAM was launched 3 times in the period 1992-95, each flight detected a clear CMB anisotropy signature. An overall analysis of the three flights yielded anisotropy detection in three band power estimates centred at $\ell$ ~ 34, 101, 407.

Boomerang and Maxima represented a major breakthough. Boomerang was launched for its first LDB flight around Antarctica in December 1998 and landed roughly 10.5 days later. An array of bolometric detectors at 90, 150, 240 and 410 GHz, cooled by a $^3$He refrigerator to 0.28K. The first results [32] were based on the data of a single detector at 150 GHz, while a more detailed analysis covering four 150GHz detectors and a larger data set was later presented [33]. Maxima, conceived as a follow-up of MAX, was a bolometric receiver cooled to 100mK by an adiabatic demagnetization refrigerator sensitive to multipoles 80<$\ell$<800 with frequency bands centred at 150, 240, and 410 GHz. The first flight in its full configuration [34] was launched in August 1998, and covered 0.3% of the sky (3200 independent pixels) during a 7-hour conventional flight. The Boomerang and Maxima power spectrum measurements were in agreement for what concerns the first-order key features: both showed clear evidence of a peak $\ell$ ~ 220 and indications of further acoustic features at higher multipoles [35]. More recently, the Archeops experiment [36] obtained accurate measurements over the multipole range 15-350 with an array of 21 bolometers in the 140-550 GHz range cooled at 100 mK with a Artic balloon flight covering 30% of the sky, and yielded accurate power spectrum data bridging between COBE and the first acoustic peak.

## WMAP, PLANCK, AND BEYOND

The results of the first-year WMAP data mark the current state of the art in the field (see [37] and references therein). The WMAP instrument is an array of differential pseudo-correlation radiometers based on HEMT amplifiers passively cooled at 90 K. The optical system is based on a pair of back-to-back off-axis telescopes which provide a highly symmetric observation scheme. The full sky is observed at five frequencies (23-94 GHz) at sub-degree scales. The far-Earth orbit around L2 and the highly redundant scanning strategy minimise the impact of systematic errors. Calibration errors are estimated <0 5%. WMAP improved the precision on the CMB dipole and quadrupole, yielding values consistent with the COBE data. The power spectrum is cosmic variance limited up to $\ell$ ~ 350 and the sensitivity is expected to improve by a factor of 2 at the end of the mission, now extended to 4 years. One of the most important results of WMAP is the observation of temperature-polarization cross power spectrum which shows the expected acoustic features and, in addition, provides evidence for a large angle correlation due to reionization at $z \sim 20$. The WMAP data have enhanced precision and robustness to the evaluation of cosmological parameters leading, in particular, to a total density $\Omega_{tot} = 1.02 \pm 0.02$ dominated by a $\Lambda$ term (see [38] for more details).

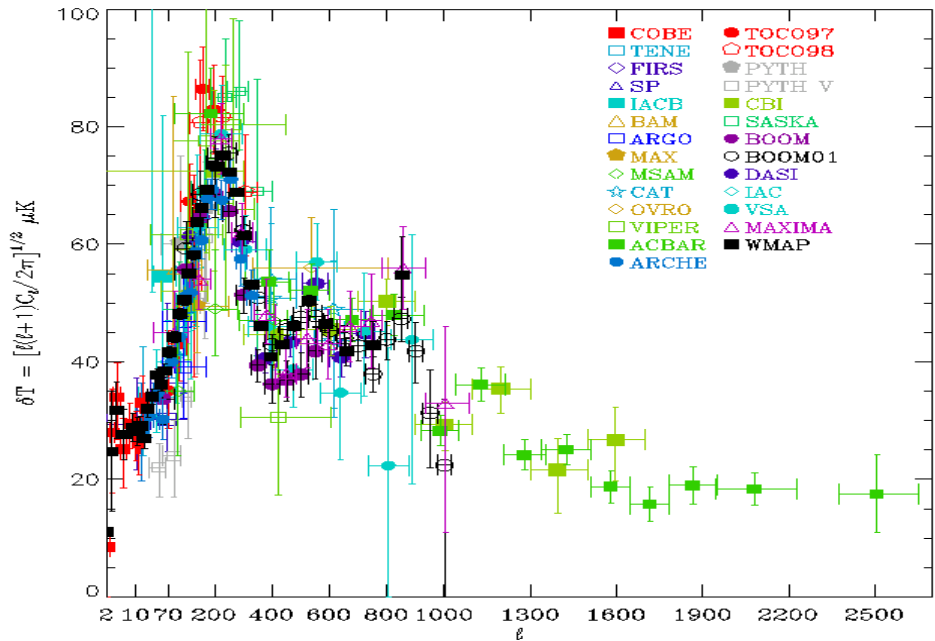

**FIGURE 1.** Present status of measurements of the CMB angular power spectrum.

The possibility of "precision cosmology" with the CMB comes from the combination of three factors. First, measurements of the CMB can be made with exquisite accuracy thanks to the rapid progress of microwave and sub-mm devices technology in recent years. Second, we have now good evidence that astrophysical foreground emission in the microwaves – mostly synchrotron, free-free and dust emission from our Galaxy, and extragalactic sources – does not prevent in principle the observation of the subtle intrinsic characteristics of the CMB. And, third, the theoretical interpretation of CMB data is relatively simple, since the features we observe in the microwave background carry information directly from epochs when all the processes were still in the linear regime.

More progress is expected in the near future. The Planck Surveyor mission, to be launched in 2007, is the third-generation CMB space observatory. Planck is expected to bring the long-lasting effort of mapping the temperaure power spectrum to completion. Although primarily designed for temperature anisotropy, Planck is capable of detecting T-E and E-mode polarisation, and of setting significant upper limits to B-modes polarisation tracing primordial gravitational waves. Exploiting both technologies traditionally used for CMB instruments (i.e. low noise HEMT amplifiers [39,40] and cryogenically cooled bolometers [41,42]) Planck will cover the whole relevant frequency range (27 to 870 GHz) with an unprecedented combination of angular resolution (<10') and sensitivity (few µK per resolution element) [43].

Rejection of systematic effects at µK level is a challenge. In some areas Planck will act as a pioneering mission rather than as a conclusive one. Two main research directions can be anticipated: precision maps of the CMB polarisation (including B-modes); and deep imaging at sub-arcminute-scales. A number of precursor sub-orbital observation programs are on-going or forthcoming in both areas. Future perspectives may include a fourth generation CMB space mission dedicated to deep polarisation observations.